\def\'#1{\ifx#1i{\accent"13\i}\else{\accent"13#1}\fi}
\def\B{{\bf B}}
\def\Egrav{E_{\rm g}}

\def\Ekin{E_{\rm k}}
\def\Emag{E_{\rm m}}
\def\Eint{E_{\rm i}}
\def\rhot{\rho_{\rm T}}
\def\u{{\bf u}}
\documentstyle[rmaaconf]{article}
 
\begin{document}

\title{CLOUD STATISTICS IN NUMERICAL SIMULATIONS OF THE ISM}

\author{Javier Ballesteros-Paredes and Enrique V\'azquez-Semadeni}
\affil{Instituto de Astronom\'{\i}a, Universidad Nacional Aut\'onoma de
M\'exico \\ Apdo. Postal 70-264, 04510 M\'{e}xico D.F., M\'{e}xico}

\begin{resumen}
Presentamos resulados preliminares de simulaciones numericas bidimensionales 
sobre el balance energ\'etico de las nubes  del medio interestelar. Mediante el uso de un algoritmo de identificaci\'on de nubes,
calculamos las energ\'ias gravitacional, interna, cin\'etica y magn\'etica
de las mismas. Encontramos que, con una dispersi\'on de aproximadamente un
orden de magnitud, la energ\'ia gravitacional en las nubes es balanceada por
las energ\'ias restantes. Adicionalmente, y con dispersiones comparables,
parece haber equipartici\'on entre las energ\'ias cin\'etica y magn\'etica.
\end{resumen}
 
\begin{abstract}
We present preliminary results on the energy budgets of clouds in
two-dimensional numerical simulations of
the interstellar medium. Using an automated cloud-identification algorithm, we
calculate the gravitational, internal, kinetic
and magnetic energies of the clouds. We find that, within a dispersion of 
roughly one order of magnitude, the gravitational energy in the clouds is 
balanced by the remaining energies. Furthermore, within the same dispersion, 
there appears to be equipartition between the kinetic and magnetic energies.
\end{abstract}


\section{INTRODUCTION}

Interstellar clouds appear to be close to virial equilibrium between the
gravitational and other forms of energy
(Larson 1981; Myers \& Goodman 1988 a,b), even though the assumption that they
are in a static equilibrium is highly questionable, as both the clouds and their
embedding medium are highly turbulent (e.g., Larson 1981; Hunter \& Fleck 1982;
Henriksen \& Turner 1984; Dickman 1985; Scalo 1987; Falgarone 1989; Fleck 
1992). Recently, V\'azquez-Semadeni, Passot \& Pouquet (1995, 
Paper I) have suggested that the apparent virialization may be due to 
nearly-virialized clouds having longer lifetimes, although the flow may not
neccessarily have a tendency towards forming virialized clouds. In order 
to test this conjecture, we have initiated a program to produce ``surveys'' of
the clouds that form in numerical simulations of the interstellar medium (ISM)
including magnetic fields (Passot, V\'azquez-Semadeni \& Pouquet, 1995, 
hereafter Paper II),
and to evaluate the various energies and terms in the virial theorem for each
cloud. 

In this paper we present preliminary results of this work. In \S 2 we 
briefly describe the numerical algorithm and define the relevant quantities. In
\S 3 we present measurements of the various energies and comparisons
that indicate rough equipartition between them. Finally, in \S 4 we summarize
and discuss the results.

\section{THE METHOD}

The two-dimensional numerical simulation from which the data are taken 
represents a square region of
1 kpc on a side in the Galactic plane, at roughly the Solar circle. Details on
the numerical method can be found in Papers I and II. The 
simulation gives the time evolution over $1.3 \times 10^8$ yr for 
all relevant physical quantities including the density, 
velocity, temperature and magnetic field, respectively measured in units of
$ \rho_0 = 1$ cm$^{-3}$, $u_0 = 11.7$ km s$^{-1}$, $T_0 = 10^4$ K and
$B_0 = 5 ~\mu$G. The gravitational, kinetic, internal and magnetic energies 
(respectively $\Egrav$, $\Ekin$, $\Eint$ and $\Emag$) are defined as
 
$$\Egrav \equiv - {1 \over 2} \int \rho \phi dV, \ \ \ \ \ \ \Ekin \equiv {1
\over 2} \int \rho \vert \u \vert ^2 dV,\ \ \ \ \ \ 
\Eint \equiv {3 \over 2} \int P dV,\ \ \ and \ \ \ \Emag \equiv {1 \over 2} \int 
\vert \B \vert ^2 dV. $$

\begin{figure}
\def\epsfsize#1#2{0.47\hsize}
\leftline{\epsffile{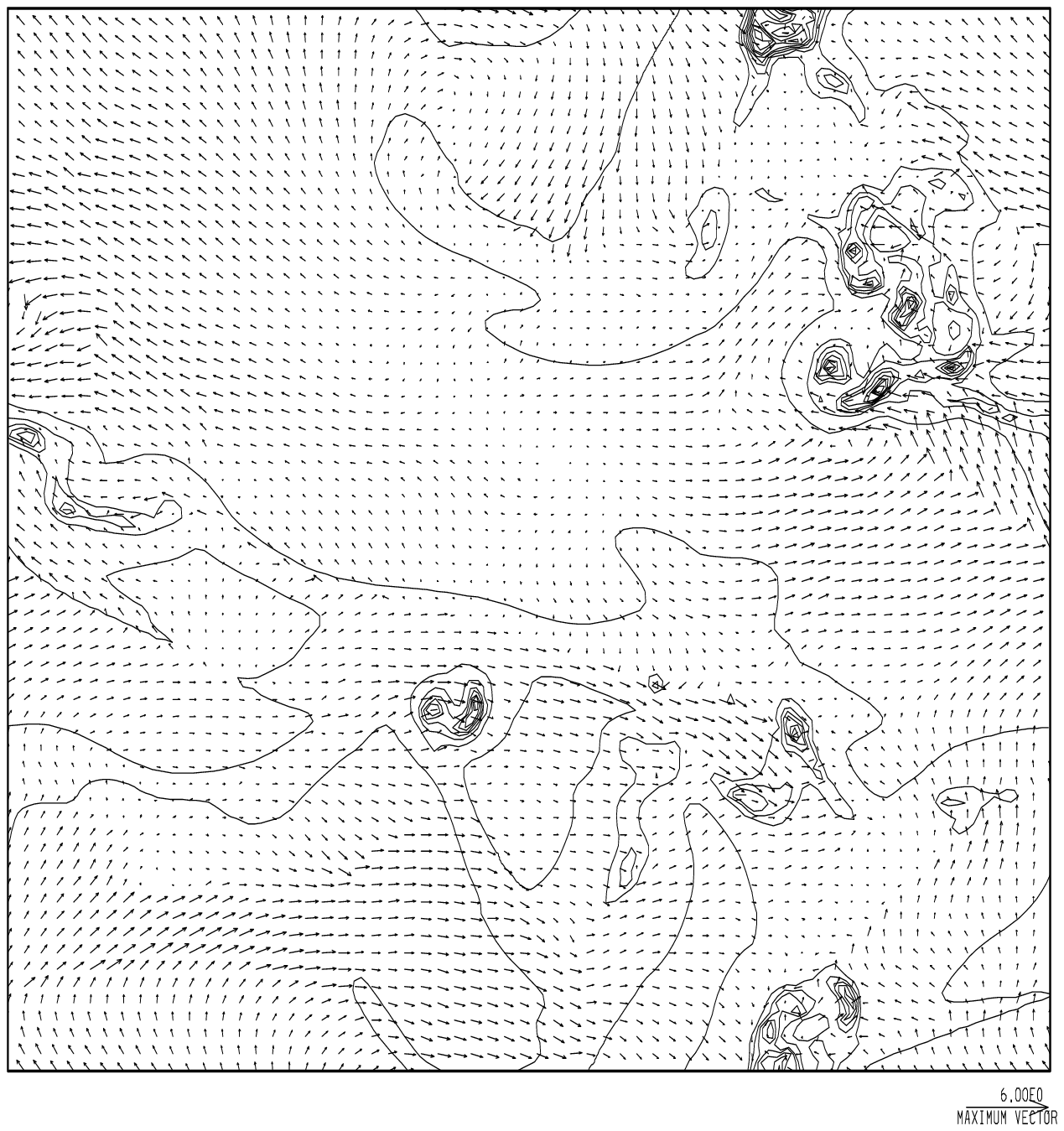}}
\vskip -1.95in
\rightline{\epsffile{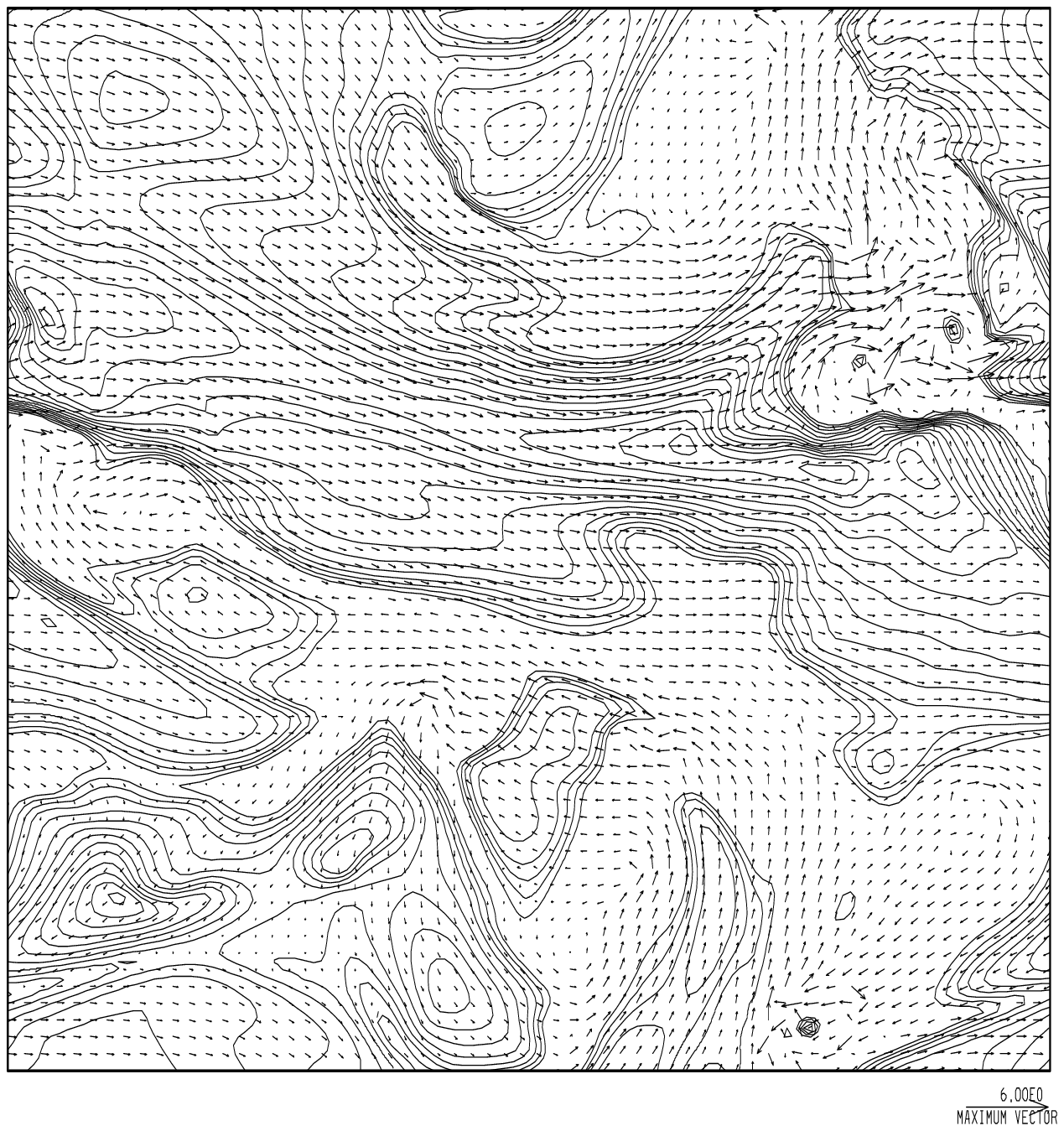}}
\caption{(Left) Density (contours) and velocity (arrows) fields of the 
numerical simulation at $t=6.6 \times 10^7$ yr into the evolution. (Right) 
Same for the temperature (contours) and magnetic field (arrows). This is a two-dimensional 
simulation with a resolution of 512 grid points per dimension.}
\label{fields}
\end{figure}

\vskip 0.25in
In Fig. \ref{fields} (left)  we show the density (contours) and the velocity
(arrows) fields at $t=6.6 \times 10^7$ yr. Similarly, Fig. \ref{fields} (right)
shows the temperature and the magnetic fields. From fig. \ref{fields}, it is 
evident that the definition of a ``cloud'' is somewhat ambiguous, as 
smaller, denser clouds are hierarchically nested within larger, less dense 
condensations, as expected for flows in which the local density probability 
distribution function is independent of the local average density and decays 
at least exponentially with the fluctuation amplitude (V\'azquez-Semadeni 
1994). In this paper we adopt a simplistic approach, and define a cloud as 
a connected set of points whose densities are larger than an arbitrary 
threshold $\rhot$. 

\begin{figure}
\def\epsfsize#1#2{0.40\hsize}
\leftline{\epsffile{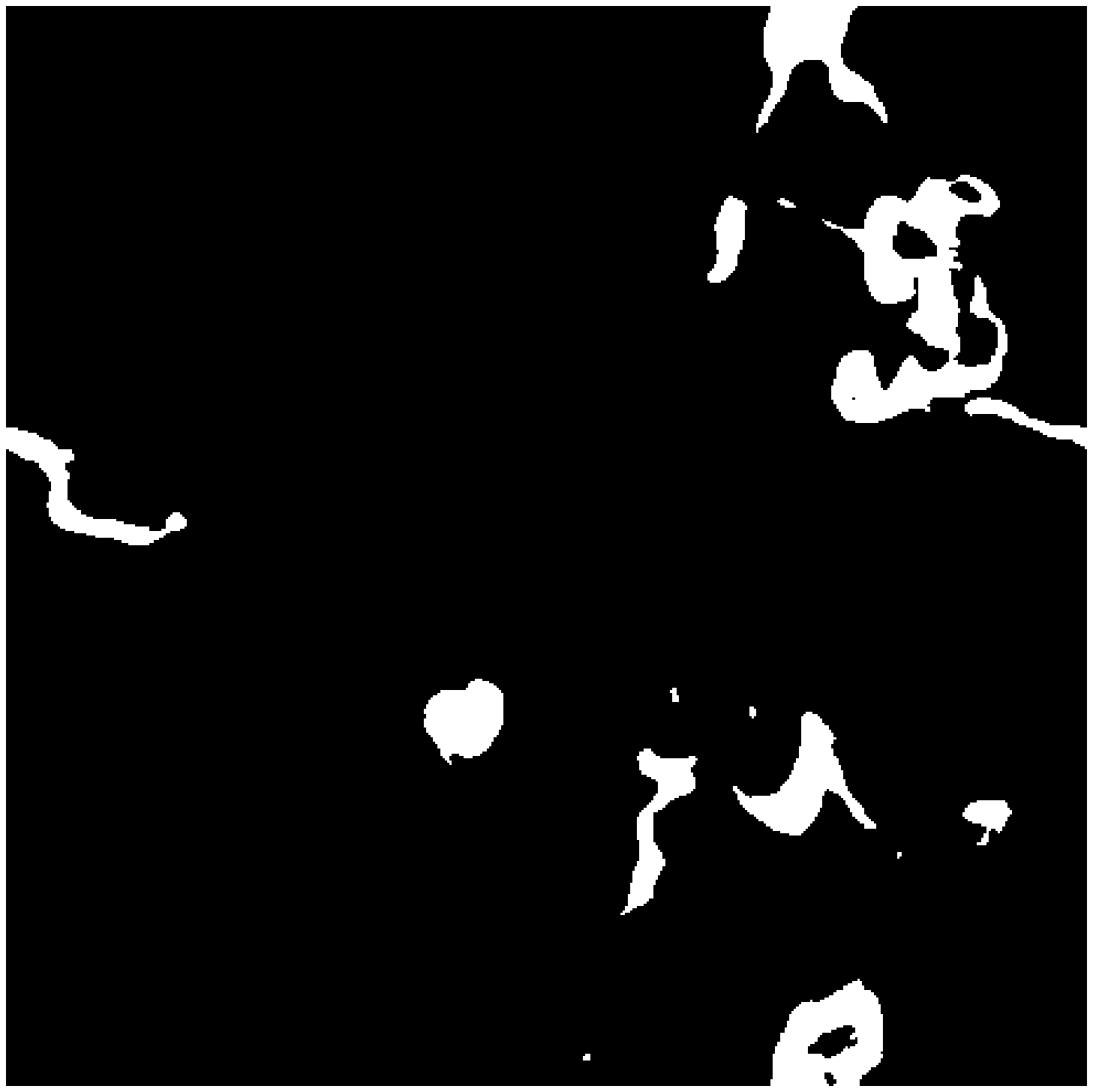}}
\vskip -3.1in
\rightline{\epsffile{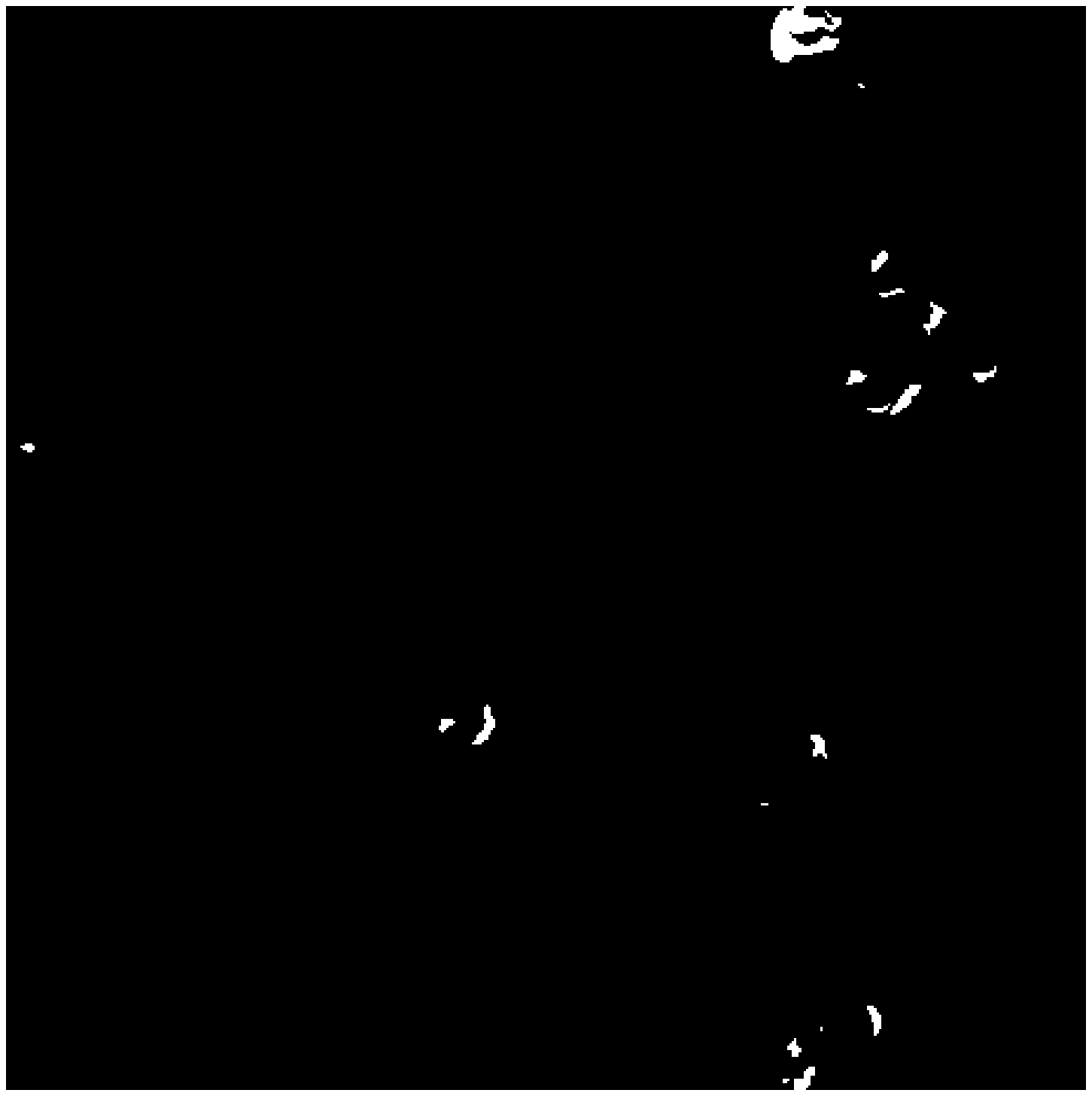}}
\caption{Masks defining the various clouds at $\rhot=4$ (left) and $\rhot=16$
(right). The integrals involved in the calculation of the various energies
are calculated as sums over the areas defined by the masks.}
\label{masks}
\end{figure}

\vskip 0.25in

With this definition, we have developed a numerical algorithm 
that identifies and labels all clouds, given $\rhot$, and produces a
``mask'' file which can be used on all the other fields (fig. \ref{masks}). 
This allows us  
to perform the above integrals inside the exact cloud perimeters and to 
evaluate the energies for each cloud. A full ``survey'' is made by taking
several values of $\rhot$, in order to include a wide range of cloud
sizes. This is equivalent to combining observations obtained 
with several different tracer molecules. 
Note also that including clouds defined through various values of $\rhot$ does
not amount to including the same cloud several times, since ``child'' clouds
may have substantially different average properties than their ``parents''.
In particular, in the discussion and figures below, we have taken 
$\rhot=$ 4 (squares), 8 (triangles) and 16 (stars) (in code units).

\section{RESULTS}

\begin{figure}
\def\epsfsize#1#2{0.35\hsize}
\centerline{\epsffile{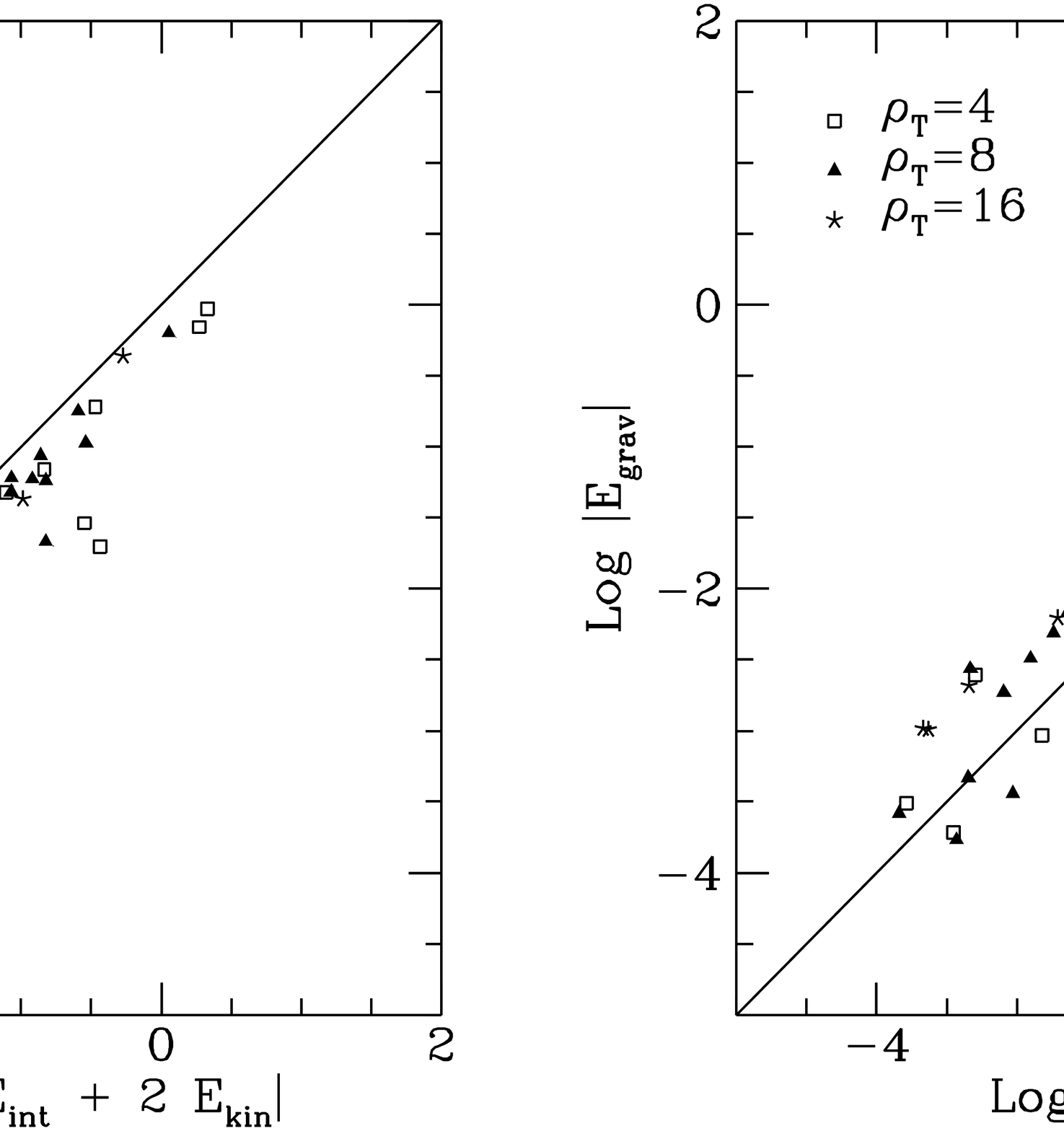}}
\caption{a) $\log{\vert{\Egrav} \vert} $ vs.\ $\log (\Emag + 2 \Eint +
2 \Ekin) $. b) $\log \vert \Egrav \vert$ vs.\ $\log (\Emag + 2 \Ekin) $}
\label{egvsetyempek}
\end{figure}

In Fig. \ref{egvsetyempek}a we show a plot of $\log{\vert{\Egrav} \vert} 
$ vs.\ $(\log \Emag +  2\Eint + 2 \Ekin)$, together with the
line $\log \vert{\Egrav} \vert = \log (\Emag + 2 \Eint + 2 \Ekin)$.
A clear correlation between the gravitational 
and the sum of the remaining energies is observed, although with a 
dispersion of roughly an order of magnitude, similar to that
found in observational studies (Myers \& Goodman 1988b; Falgarone
et al. 1992). Furthermore, it appears that the gravitational energy is
systematically too low, and thus, according to this plot, most
clouds in the simulation are not gravitationally bound.

However, it is well known (e.g., Shu 1990) that if the pressure is nearly
uniform throughout the flow, then the contribution of the internal energy  of
the cloud is nearly balanced by the external pressure acting on the boundary of
the cloud. In the simulation, the typical pressure contrast between clouds and
the intercloud medium is $\sim 5,$ except in regions of star formation. The 
exact contribution of the external pressure requires an integral
over the cloud's boundary, which will be discussed in a future paper. 
In this preliminary report, we show in 
fig. \ref{egvsetyempek}b a plot of $\log \vert{\Egrav} 
\vert$ vs.\ $\log (\Emag + 2 \Ekin)$, which corresponds to the case in
which the surface pressure term exactly
balances $2 \Eint$. The actual situation must lie between
the limiting cases depicted in Figs. \ref{egvsetyempek}a and 3b.

\begin{figure}
\def\epsfsize#1#2{0.35\hsize}
\centerline{\epsffile{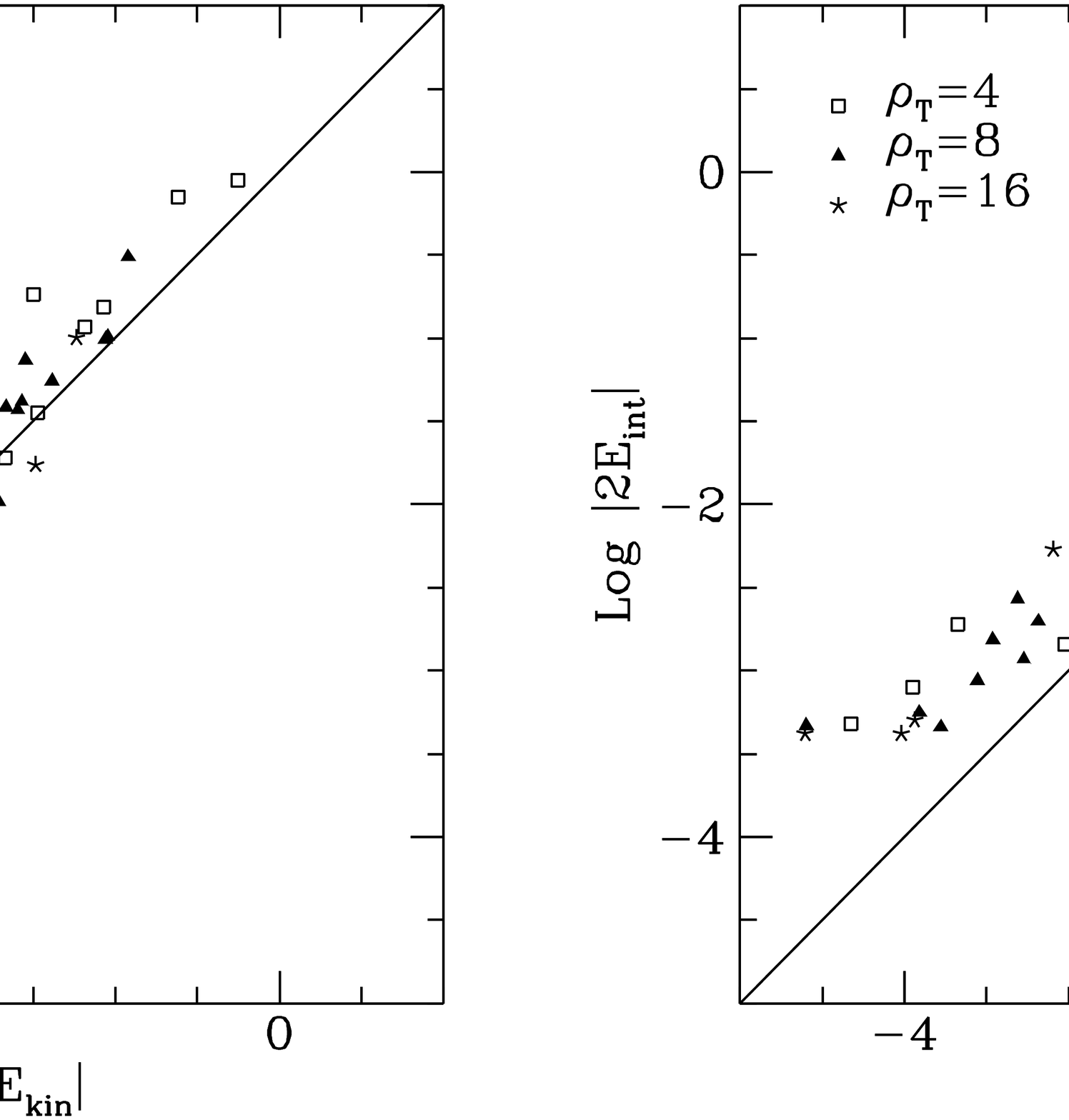}}
\caption{a) $\log{\vert{\Egrav} \vert} $ vs.\ $\log (\Emag + 2 \Eint +
2 \Ekin) $. b) $\log \vert \Egrav \vert$ vs.\ $\log (\Emag + 2 \Ekin) $}
\label{em.vs.ek.ei.vs.ek}
\end{figure}

In order to search for possible equipartition among different forms of energy,
in fig. \ref{em.vs.ek.ei.vs.ek}a we show a plot of $\log ({\Emag} )$ vs.\ $\log ({2 \Ekin} 
)$, and in fig. \ref{em.vs.ek.ei.vs.ek}b we show a plot of $\log ({2 \Eint} )$ vs.\ $\log
({2 \Ekin} )$.
A tendency towards equipartition between the turbulent kinetic energy and the
magnetic and internal energies is apparent,
again with a typical dispersion of roughly an order of magnitude. Note that in
fig. \ref{em.vs.ek.ei.vs.ek}b a deviation from equipartition seen
at low values of the energies is probably an artifact of the
simulation, because at small scales the velocity is damped by dissipation.

\section{CONCLUSIONS}

In this paper we have introduced a simple algorithm for identifying all clouds
in  numerical simulations of the ISM and presented preliminary statistics over
the cloud sample. The gravitational energy appears to be in rough
balance with the remaining forms of energy, although with non-negligible
dispersion. This result is similar to those obtained from observations (e.g., 
Larson 1981; Myers \& Goodman 1988a,b), and has been taken in those works 
to be indicative of near-virialization in clouds. However, we emphasize that 
the surface terms in the virial theorem may have an important, if not 
decisive, contribution to the overall virial balance of the clouds, as 
already suggested by the role of the thermal pressure in the data presented 
here.

The fact that there seems to be a trend for clouds to exhibit balance between
gravity and its opposing agents suggests that the flow may indeed have
a tendency towards producing nearly-virial clouds, in contradiction with the
conjecture in Paper I. However, the relatively small dispersion about this
balance may be an artifact of the low contrast in $\rhot$ used here due to
numerical difficulties. This selects against small, low-density clouds in which
gravity is sub-dominant. This difficulty, plus the contribution of the surface 
terms, the necessary modifications to the virial theorem in two-dimensions, 
and the longevity of clouds as a function of their closeness to virial 
balance, all need to be assesed before a conclusive answer can be given. 
Work is currently in progress to address these issues.

\end{document}